\documentstyle[twoside]{article}
\input{epsf} 
\newcommand\jhep[3]{JHEP {\bf #1}, #3 (#2)} 
\newcommand\npb[3]{Nucl.\ Phys.\ B {\bf #1}, #3 (#2)} 
\newcommand\plb[3]{Phys.\ Lett.\ B {\bf #1}, #3 (#2)} 
\newcommand\Prd[3]{Phys.\ Rev.\ D {\bf #1}, #3 (#2)}
\newcommand\Prl[3]{Phys.\ Rev.\ Lett.\ {\bf #1}, #3 (#2)}

\catcode`\@=11
\long\def\@makefntext#1{
\protect\noindent \hbox to 3.2pt {\hskip-.9pt  
$^{{\eightrm\@thefnmark}}$\hfil}#1\hfill}		

\def\@makefnmark{\hbox to 0pt{$^{\@thefnmark}$\hss}}	
	
\def\ps@myheadings{\let\@mkboth\@gobbletwo
\def\@oddhead{\hbox{}
\rightmark\hfil\eightrm\thepage}   
\def\@oddfoot{}\def\@evenhead{\eightrm\thepage\hfil
\leftmark\hbox{}}\def\@evenfoot{}
\def\sectionmark##1{}\def\subsectionmark##1{}}

\oddsidemargin=\evensidemargin
\newcounter{sectionc}\newcounter{subsectionc}\newcounter{subsubsectionc}
\renewcommand{\section}[1] {\vspace{12pt}\addtocounter{sectionc}{1} 
\setcounter{subsectionc}{0}\setcounter{subsubsectionc}{0}\noindent 
	{\tenbf\thesectionc. #1}\par\vspace{5pt}}
\renewcommand{\subsection}[1] {\vspace{12pt}\addtocounter{subsectionc}{1} 
	\setcounter{subsubsectionc}{0}\noindent 
	{\bf\thesectionc.\thesubsectionc. {\kern1pt \bfit #1}}\par\vspace{5pt}}
\renewcommand{\subsubsection}[1] {\vspace{12pt}\addtocounter{subsubsectionc}{1}
	\noindent{\tenrm\thesectionc.\thesubsectionc.\thesubsubsectionc.
	{\kern1pt \tenit #1}}\par\vspace{5pt}}
\newcommand{\nonumsection}[1] {\vspace{12pt}\noindent{\tenbf #1}
	\par\vspace{5pt}}

\newcounter{appendixc}
\newcounter{subappendixc}[appendixc]
\newcounter{subsubappendixc}[subappendixc]
\renewcommand{\thesubappendixc}{\Alph{appendixc}.\arabic{subappendixc}}
\renewcommand{\thesubsubappendixc}
	{\Alph{appendixc}.\arabic{subappendixc}.\arabic{subsubappendixc}}

\renewcommand{\appendix}[1] {\vspace{12pt}
        \refstepcounter{appendixc}
        \setcounter{figure}{0}
        \setcounter{table}{0}
        \setcounter{lemma}{0}
        \setcounter{theorem}{0}
        \setcounter{corollary}{0}
        \setcounter{definition}{0}
        \setcounter{equation}{0}
        \renewcommand{\thefigure}{\Alph{appendixc}.\arabic{figure}}
        \renewcommand{\thetable}{\Alph{appendixc}.\arabic{table}}
        \renewcommand{\theappendixc}{\Alph{appendixc}}
        \renewcommand{\thelemma}{\Alph{appendixc}.\arabic{lemma}}
        \renewcommand{\thetheorem}{\Alph{appendixc}.\arabic{theorem}}
        \renewcommand{\thedefinition}{\Alph{appendixc}.\arabic{definition}}
        \renewcommand{\thecorollary}{\Alph{appendixc}.\arabic{corollary}}
        \renewcommand{\theequation}{\Alph{appendixc}.\arabic{equation}}
        \noindent{\tenbf Appendix \theappendixc #1}\par\vspace{5pt}}
\newcommand{\subappendix}[1] {\vspace{12pt}
        \refstepcounter{subappendixc}
        \noindent{\bf Appendix \thesubappendixc. {\kern1pt \bfit #1}}
	\par\vspace{5pt}}
\newcommand{\subsubappendix}[1] {\vspace{12pt}
        \refstepcounter{subsubappendixc}
        \noindent{\rm Appendix \thesubsubappendixc. {\kern1pt \tenit #1}}
	\par\vspace{5pt}}

\newcommand{\textlineskip}{\baselineskip=13pt}
\newcommand{\smalllineskip}{\baselineskip=10pt}

\def\eightcirc{
\begin{picture}(0,0)
\put(4.4,1.8){\circle{6.5}}
\end{picture}}
\def\eightcopyright{\eightcirc\kern2.7pt\hbox{\eightrm c}} 

\newcommand{\copyrightheading}[1]
	{\vspace*{-2.5cm}\smalllineskip{\flushleft
	{\footnotesize International Journal of Modern Physics A, #1}\\
	{\footnotesize $\eightcopyright$\, World Scientific Publishing
	 Company}\\
	 }}

\def\abstracts#1#2#3{{
	\centering{\begin{minipage}{4.5in}\baselineskip=10pt\footnotesize
	\parindent=0pt #1\par 
	\parindent=15pt #2\par
	\parindent=15pt #3
	\end{minipage}}\par}} 

\renewenvironment{thebibliography}[1]
	{\frenchspacing
	 \ninerm\baselineskip=11pt
	 \begin{list}{\arabic{enumi}.}
	{\usecounter{enumi}\setlength{\parsep}{0pt}
	 \setlength{\leftmargin 12.7pt}{\rightmargin 0pt} 
	 \setlength{\itemsep}{0pt} \settowidth
	{\labelwidth}{#1.}\sloppy}}{\end{list}}

\newcounter{itemlistc}
\newcounter{romanlistc}
\newcounter{alphlistc}
\newcounter{arabiclistc}

\newcommand{\fcaption}[1]{
        \refstepcounter{figure}
        \setbox\@tempboxa = \hbox{\footnotesize Fig.~\thefigure. #1}
        \ifdim \wd\@tempboxa > 5in
           {\begin{center}
        \parbox{5in}{\footnotesize\smalllineskip Fig.~\thefigure. #1}
            \end{center}}
        \else
             {\begin{center}
             {\footnotesize Fig.~\thefigure. #1}
              \end{center}}
        \fi}

\newcommand{\tcaption}[1]{
        \refstepcounter{table}
        \setbox\@tempboxa = \hbox{\footnotesize Table~\thetable. #1}
        \ifdim \wd\@tempboxa > 5in
           {\begin{center}
        \parbox{5in}{\footnotesize\smalllineskip Table~\thetable. #1}
            \end{center}}
        \else
             {\begin{center}
             {\footnotesize Table~\thetable. #1}
              \end{center}}
        \fi}

\def\@citex[#1]#2{\if@filesw\immediate\write\@auxout
	{\string\citation{#2}}\fi
\def\@citea{}\@cite{\@for\@citeb:=#2\do
	{\@citea\def\@citea{,}\@ifundefined
	{b@\@citeb}{{\bf ?}\@warning
	{Citation `\@citeb' on page \thepage \space undefined}}
	{\csname b@\@citeb\endcsname}}}{#1}}

\newif\if@cghi
\def\cite{\@cghitrue\@ifnextchar [{\@tempswatrue
	\@citex}{\@tempswafalse\@citex[]}}
\def\citelow{\@cghifalse\@ifnextchar [{\@tempswatrue
	\@citex}{\@tempswafalse\@citex[]}}
\def\@cite#1#2{{$\null^{#1}$\if@tempswa\typeout
	{IJCGA warning: optional citation argument 
	ignored: `#2'} \fi}}

\def\pmb#1{\setbox0=\hbox{#1}
	\kern-.025em\copy0\kern-\wd0
	\kern.05em\copy0\kern-\wd0
	\kern-.025em\raise.0433em\box0}

\def\fnt#1#2{\footnotetext{\kern-.3em
	{$^{\mbox{\scriptsize #1}}$}{#2}}}

\def\fpage#1{\begingroup
\voffset=.3in
\thispagestyle{empty}\begin{table}[b]\centerline{\footnotesize #1}
	\end{table}\endgroup}

\def\runninghead#1#2{\pagestyle{myheadings}
\markboth{{\protect\footnotesize\it{\quad #1}}\hfill}
{\hfill{\protect\footnotesize\it{#2\quad}}}}
\font\tenrm=cmr10
\font\tenit=cmti10 
\font\tenbf=cmbx10
\font\bfit=cmbxti10 at 10pt
\font\ninerm=cmr9

\font\eightrm=cmr8

\def\qed{\hbox{${\vcenter{\vbox{			
   \hrule height 0.4pt\hbox{\vrule width 0.4pt height 6pt
   \kern5pt\vrule width 0.4pt}\hrule height 0.4pt}}}$}}

\begin{document}

\runninghead{Enhanced Gluino-Box Contribution to $\epsilon'/\epsilon$ 
in SUSY}
{Enhanced Gluino-Box Contribution to $\epsilon'/\epsilon$ in SUSY} 

\normalsize\textlineskip
\thispagestyle{empty}
\setcounter{page}{1}

\copyrightheading{}			

\vspace*{0.88truein}

\fpage{1}
\centerline{\bf\boldmath ENHANCED GLUINO BOX CONTRIBUTION TO 
$\epsilon'/\epsilon$ IN SUSY\unboldmath}
\vspace*{0.37truein}
\centerline{\footnotesize MATTHIAS NEUBERT}
\vspace*{0.015truein}
\centerline{\footnotesize\it Newman Laboratory of Nuclear Studies, 
Cornell University}
\baselineskip=10pt
\centerline{\footnotesize\it Ithaca, New York 14853, U.S.A.}

\vspace*{0.21truein}
\abstracts{
We show that in supersymmetric extensions of the Standard Model 
gluino box diagrams can yield a large CP-violating $\Delta I=\frac32$
contribution to $s\to d\bar q q$ flavor-changing neutral current 
processes, which feeds into the $I=2$ isospin amplitude in 
$K\to\pi\pi$ decays. This contribution only requires moderate
mass splitting between the right-handed squarks $\tilde u_R$ and 
$\tilde d_R$, and persists for squark masses of order 1\,TeV. 
Taking into account current bounds on $\mbox{Im}\,\delta_{sd}^{LL}$ 
from $K$--$\bar K$ mixing, the resulting contribution to 
$\epsilon'/\epsilon$ could naturally be an order of magnitude larger 
than the measured value.}{}{}

\textlineskip			
\vspace*{12pt}			


The recent confirmation of direct CP violation in $K\to\pi\pi$ 
decays is an important step in testing the 
Cabibbo--Kobayashi--Maskawa (CKM) mechanism for CP violation in 
the Standard Model. Combining the recent measurements by the KTeV 
and NA48 experiments with earlier results from NA31 and 
E731 gives $\mbox{Re}\,(\epsilon'/\epsilon)
=(2.12\pm 0.46)\times 10^{-3}$. This value tends to be higher than 
theoretical predictions in the Standard Model, which center below 
or around $1\times 10^{-3}$. Unfortunately, it is 
difficult to gauge the accuracy of these predictions, because they 
depend on hadronic matrix elements which at present cannot be 
computed from first principles. A Standard-Model explanation of 
$\epsilon'/\epsilon$ can therefore not be excluded. Nevertheless, 
it is interesting to ask how large $\epsilon'/\epsilon$ could be 
in extensions of the Standard Model.

In the context of supersymmetric models, it has been known for some 
time that it is possible to obtain a large contribution to 
$\epsilon'/\epsilon$ via the $\Delta I=\frac12$ chromomagnetic 
dipole operator without violating constraints from $K$--$\bar K$ 
mixing.\cite{Gabb} It has recently been pointed out that this 
mechanism can naturally be realized in various supersymmetric 
scenarios.\cite{SUSY} Here we discuss a new mechanism 
involving a supersymmetric contribution to $\epsilon'/\epsilon$ 
induced by $\Delta I=\frac32$ penguin operators.\cite{us} 
These operators 
are potentially important because their effect is enhanced by the 
$\Delta I=\frac12$ selection rule. Unlike recent proposals, which 
involve left-right down-squark mass insertions, our effect relies 
on the left-left insertion $\delta_{sd}^{LL}$ and requires (moderate) 
isospin violation in the right-handed squark sector. 

The ratio $\epsilon'/\epsilon$ parameterizing the strength of direct 
CP violation in $K\to\pi\pi$ decays can be expressed as
\begin{equation}\label{eps}
   \frac{\epsilon'}{\epsilon}
   \simeq \frac{\omega}{\sqrt2\,|\epsilon|} \left( 
   \frac{\mbox{Im}\,A_2}{\mbox{Re}\,A_2}
   - \frac{\mbox{Im}\,A_0}{\mbox{Re}\,A_0} \right) ,
\end{equation}
where $A_I$ are the isospin amplitudes for the decays $K^0\to
(\pi\pi)_I$, and the ratio 
$\omega=\mbox{Re}\,A_2/\mbox{Re}\,A_0\approx 0.045$ signals the 
strong enhancement of $\Delta I=\frac12$ 
transitions over those with $\Delta I=\frac32$. 
In the Standard Model, the amplitudes $A_I$ receive small, 
CP-violating contributions via the ratio 
$(V_{ts}^* V_{td})/(V_{us}^* V_{ud})$ of CKM matrix elements, which 
enters through the interference of the $s\to u\bar u d$ tree 
diagram with penguin diagrams involving the exchange of a virtual 
top quark. According to (\ref{eps}), contributions to 
$\epsilon'/\epsilon$ due to the $\Delta I=\frac32$ amplitude 
$\mbox{Im}\,A_2$ are enhanced relative to those due to the 
$\Delta I=\frac12$ amplitude $\mbox{Im}\,A_0$ by a factor of 
$\omega^{-1}\approx 22$. However, in the Standard Model the dominant 
CP-violating contributions to $\epsilon'/\epsilon$ are due to QCD 
penguin operators, which only contribute to $A_0$. Penguin 
contributions to $A_2$ arise through electroweak interactions and 
are suppressed by a power of $\alpha$. Their effects on 
$\epsilon'/\epsilon$ are small and of the same order as 
isospin-violating effects such as $\pi^0$--$\eta$--$\eta'$ mixing.

In supersymmetric extensions of the Standard 
Model potentially large, CP-violating contributions can arise from 
flavor-changing {\em strong-interaction\/} processes induced by 
gluino box diagrams. Whereas in the limit of exact isospin symmetry
in the squark sector these graphs only induce $\Delta I=\frac12$ 
operators at low energies, in the presence of even moderate up--down 
squark mass splitting they can lead to operators with large 
$\Delta I=\frac32$ components. In the terminology of the standard 
effective weak Hamiltonian, this implies that the supersymmetric 
contributions to the Wilson coefficients of QCD and electroweak 
penguin operators can be of the same order. Specifically, both sets 
of coefficients scale like $\alpha_s^2/\widetilde m^2$ with 
$\widetilde m$ a generic supersymmetric mass, compared with 
$\alpha_s\alpha_W/m_W^2$ and $\alpha\alpha_W/m_W^2$, respectively, 
in the Standard Model. These contributions can be much larger than 
the electroweak penguins of the Standard Model even for 
supersymmetric masses of order 1\,TeV. 

The relevant $\Delta S=1$ gluino box diagrams lead to 
the effective Hamiltonian 
\[
   {\cal H}_{\rm eff} = \frac{G_F}{\sqrt2}
   \sum_{i=1}^4 \left[ c_i^q(\mu)\,Q_i^q(\mu)
   + \widetilde c_i^q(\mu)\,\widetilde Q_i^q(\mu) \right] 
   + \mbox{h.c.} \,,
\]
where 
\begin{eqnarray}
   Q_1^q &=& (\bar d_\alpha s_\alpha)_{V-A}\,
    (\bar q_\beta q_\beta)_{V+A} \,, \qquad
    Q_2^q = (\bar d_\alpha s_\beta)_{V-A}\,
    (\bar q_\beta q_\alpha)_{V+A} \,, \nonumber\\
   Q_3^q &=& (\bar d_\alpha s_\alpha)_{V-A}\,
    (\bar q_\beta q_\beta)_{V-A} \,, \qquad
    Q_4^q = (\bar d_\alpha s_\beta)_{V-A}\,
    (\bar q_\beta q_\alpha)_{V-A} \nonumber
\end{eqnarray}
are local four-quark operators, $\widetilde Q_i^q$ are operators of 
opposite
chirally obtained by interchanging $V-A\leftrightarrow V+A$, and a 
summation over $q=u,d,\dots$ and over color indices $\alpha,\beta$ 
is implied. In the calculation of the coefficient functions we use
the mass insertion approximation, in which case the 
gluino--quark--squark couplings are flavor diagonal. Flavor mixing
is due to small deviations from squark-mass degeneracy and is 
parameterized by dimensionless quantities $\delta_{ij}^{AB}$, where 
$i,j$ are squark flavor indices and $A,B$ refer to the chiralities 
of the corresponding quarks. In general, 
these mass insertions can carry new CP-violating phases. 
Contributions involving left-right squark mixing are neglected, 
since they are quadratic in small mass insertion parameters, i.e., 
proportional to $\delta_{sd}^{LR}\,\delta_{qq}^{LR}$. We define 
the dimensionless mass ratios 
$x_u^{L,R}=(m_{\tilde u_{L,R}}/m_{\tilde g})^2$
and $x_d^{L,R}=(m_{\tilde d_{L,R}}/m_{\tilde g})^2$,
where $m_{\tilde u_{L,R}}$ and $m_{\tilde d_{L,R}}$ denote the 
average left- or right-handed squark masses in the up and down 
sector, respectively. SU(2)$_L$ gauge symmetry implies that the 
mass splitting between the left-handed up- and down-squarks
must be tiny; however, we will not assume such a degeneracy
between the right-handed squarks. 

It is straightforward to relate the quantities $c_i^q$ to the 
Wilson coefficients appearing in the effective weak Hamiltonian 
of the Standard Model. 
The supersymmetric contributions to the electroweak penguin 
coefficients vanish in the limit of universal squark masses. 
However, for moderate up--down squark mass splitting the 
differences $\Delta c_i\equiv c_i^u-c_i^d$ are of the same order as 
the coefficients $c_i^q$ themselves. In this case gluino box 
contributions to QCD and electroweak penguin operators are of 
similar magnitude. Because the electroweak 
penguin operators contain $\Delta I=\frac32$ components their 
contributions to $\epsilon'/\epsilon$ are strongly enhanced and 
thus are expected to be an order of magnitude larger than the 
contributions from the QCD penguin operators. In this talk, we 
focus only on these enhanced contributions. Since the mass 
splitting between the left-handed $\tilde u_L$ and $\tilde d_L$ 
squarks is tiny, we can safely neglect the coefficients 
$\Delta c_{3,4}$ and $\Delta\widetilde c_{1,2}$ in our analysis.
Finally, in estimating the supersymmetric contribution to 
$\epsilon'/\epsilon$ we focus only on the $(V-A)\otimes (V+A)$ 
operators associated with the coefficients $\Delta c_1$ and 
$\Delta c_2$, because their matrix elements are chirally enhanced 
with respect to those of the $(V+A)\otimes(V+A)$ operators. 

Table~\ref{tab:1} shows the imaginary parts of the coefficients 
$\Delta c_{1,2}$ obtained for the illustrative choice 
$\widetilde m=m_{\tilde g}=m_{\tilde d_L}=m_{\tilde d_R}=500$\,GeV 
and three (larger) values of $m_{\tilde u_R}$. 
For fixed ratios of the supersymmetric masses the 
values of the coefficients scale like $\widetilde m^{-2}$, i.e., 
significantly larger values could be obtained for smaller masses. 
For comparison, the last column contains the imaginary parts of 
$\Delta c_{1,2}$ in the Standard Model. We observe that for 
supersymmetric masses of order 
500\,GeV, and for a mass insertion parameter 
$\mbox{Im}\,\delta_{sd}^{LL}$ of order a few times $10^{-3}$ 
(see below), the imaginary part of $\Delta c_2$ can be 
significantly larger than that of the corresponding 
coefficient in the Standard Model, which is proportional to 
$C_8$. This is interesting, since even in the Standard Model the 
contribution of $C_8$ to $\epsilon'/\epsilon$ is 
significant.

\begin{table}
\caption{\label{tab:1} 
Imaginary parts of the coefficients $\Delta c_{1,2}(m_c)$ in units 
of $10^{-4}\,\mbox{Im}\,\delta_{sd}^{LL}$, for gluino and down-squark 
masses of 500\,GeV and different values of $m_{\tilde u_R}$. The last 
column shows the corresponding values in the Standard Model.}
\vspace{0.4cm}
\begin{center}
\begin{tabular}{l|rrr|r}
\hline
$m_{\tilde u_R}$\,[GeV] & 750 & 1000 & 1500 & SM \\
\hline
Im$\Delta c_1(m_c)$ & $-0.05$ & $-0.08$ & $-0.12$
 & $0.42\times 10^{-7}$ \\ 
Im$\Delta c_2(m_c)$ & 2.12 & 3.19 & 4.16& $-1.90\times 10^{-7}$ \\
\hline
\end{tabular}
\end{center}
\vspace{-0.4cm}
\end{table}

\begin{figure}
\begin{center}
\epsfxsize=11.8cm
\epsffile{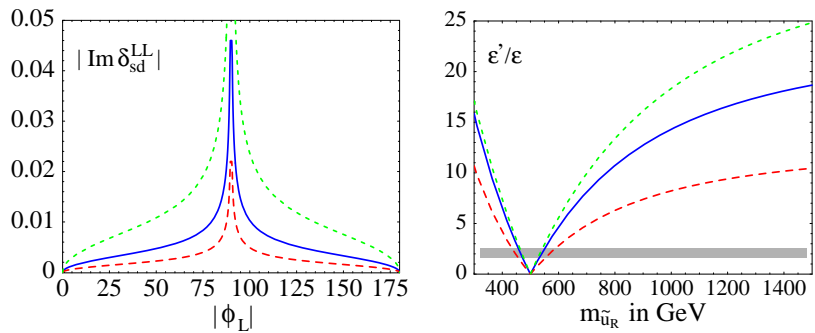}
\end{center}
\vspace{-0.6cm}
\caption{\label{fig:bounds}
Left:
Upper bound on $|\mbox{Im}\,\delta_{sd}^{LL}|$ versus the weak 
phase $|\phi_L|$ (in degrees) for $m_{\tilde d_L}=500$\,GeV and 
$(m_{\tilde g}/m_{\tilde d_L})^2=1$ (solid), 0.3 (dashed) and 4
(short-dashed).
Right: 
Supersymmetric contribution to $|\epsilon'/\epsilon|$ (in units
of $10^{-3}$) versus $m_{\tilde u_R}$, for $m_s(m_c)=130$\,MeV,
$B_8^{(2)}(m_c)=1$, $m_{\tilde d_L}=m_{\tilde d_R}=500$\,GeV, 
and $(m_{\tilde g}/m_{\tilde d_L})^2=1$ (solid), 0.3 (dashed) 
and 4 (short-dashed). The values of 
$|\mbox{Im}\,\delta_{sd}^{LL}|$ corresponding to the three
curves are 0.011, 0.005 and 0.027, respectively (see text). The 
band shows the average experimental value.}
\end{figure}

The penguin operators contribute to the imaginary part of the 
isospin amplitude $A_2$. The real part is, to an excellent 
approximation, given by the matrix elements of the standard 
current--current operators in the effective weak Hamiltonian. 
Evaluating the matrix elements of the four-quark operators in the 
factorization approximation, and parameterizing nonfactorizable 
corrections by a hadronic parameter $B_8^{(2)}$, we 
find for the supersymmetric
$\Delta I=\frac32$ contribution to $\epsilon'/\epsilon$
\[
   \frac{\epsilon'}{\epsilon} \approx 19.2 
   \left[ \frac{500\,\mbox{GeV}}{m_{\tilde g}} \right]^2
   \left[ \frac{\alpha_s(\widetilde m)}{0.096} 
   \right]^\frac{34}{21}
   \left[ \frac{130\,\mbox{MeV}}{m_s(m_c)} \right]^2
   B_8^{(2)}(m_c)\,X(x_d^L,x_u^R,x_d^R)\,
   \mbox{Im}\,\delta_{sd}^{LL} \,,
\]
where $X(x,y,z)$ is a known function of SUSY mass ratios.
The existence of this contribution requires a new CP-violating
phase $\phi_L$ defined by $\mbox{Im}\,\delta_{sd}^{LL}\equiv
|\delta_{sd}^{LL}|\sin\phi_L$. The measured values of 
$\Delta m_K$ and $\epsilon$ in $K$--$\bar K$ mixing give bounds 
on $\mbox{Re}\,(\delta_{sd}^{LL})^2$ and 
$\mbox{Im}\,(\delta_{sd}^{LL})^2$, respectively, which can be 
combined to obtain a bound on $\mbox{Im}\,\delta_{sd}^{LL}$ as a 
function of $\phi_L$. Using the most recent analysis of 
supersymmetric contributions to $K$--$\bar K$ mixing in 
Ref.~\cite{long}, we show in the left-hand plot in 
Fig.~\ref{fig:bounds} the results 
obtained for $m_{\tilde d_L}=500$\,GeV and three choices of 
$m_{\tilde g}$. It is evident that the bound on 
$\mbox{Im}\,\delta_{sd}^{LL}$ depends strongly on the precise 
value of $\phi_L$. To address the issue of how large a 
supersymmetric contribution to $\epsilon'/\epsilon$ one can 
reasonably expect from our mechanism,
it appears unnatural to take the absolute maximum of the bound 
given the peaked nature of the curves. We thus 
evaluate our result taking for 
$\mbox{Im}\,\delta_{sd}^{LL}$ one quarter of the maximal allowed 
values. In this way, we obtain a plausible upper bound on the 
SUSY contribution to $\epsilon'/\epsilon$, which does not require
fine-tuning. 

Our results for $|\epsilon'/\epsilon|$ are shown in the right-hand 
plot in
Fig.~\ref{fig:bounds} as a function of $m_{\tilde u_R}$ for the 
case $m_{\tilde d_L}=m_{\tilde d_R}=500$\,GeV and the same three
values of $m_{\tilde g}$ considered in the previous figure. The 
choice $m_{\tilde d_L}=m_{\tilde d_R}$ is made for simplicity 
only and does not affect our conclusions in a qualitative way.
Except for the special case of highly degenerate right-handed 
up- and down-squark masses, the $\Delta I=\frac32$ gluino 
box-diagram contribution to $\epsilon'/\epsilon$ can by far 
exceed the experimental result, even taking into account the 
bounds from $\Delta m_K$ and $\epsilon$. Indeed, even for 
moderate splitting the figure implies substantially 
stronger bounds on $|\mbox{Im}\,\delta_{sd}^{LL}|$ than those 
obtained from $K$--$\bar K$ mixing. This finding is in contrast 
to the commonly held view that supersymmetric contributions to 
the electroweak penguin operators have a negligible impact on 
$\epsilon'/\epsilon$. In this context, it is worth noting that 
a large mass splitting between $\tilde u_R$ and $\tilde d_R$ 
can be obtained, e.g., in GUT theories without SU(2)$_R$ 
symmetry and with hypercharge embedded in the unified gauge 
group, without encountering difficulties with naturalness.\cite{Gian}

The allowed contribution to $\epsilon'/\epsilon$ 
increases with the gluino mass (for fixed 
squark masses) because the $K$--$\bar K$ bounds become weaker in 
this case. If all supersymmetric masses are rescaled by a common 
factor $\xi$, and the bounds from $K$--$\bar K$ mixing are 
rescaled accordingly, the values for $\epsilon'/\epsilon$ scale 
like $1/\xi$. Even for larger 
squark masses of order 1\,TeV the new contribution to 
$\epsilon'/\epsilon$ can exceed the experimental value.

In the above discussion we have
made no assumption regarding the mass insertion parameter 
$\mbox{Im}\,\delta_{sd}^{RR}\equiv|\delta_{sd}^{RR}|\sin\phi_R$
for right-handed squarks. In models where $|\delta_{sd}^{RR}|$ 
is not highly suppressed relative to $|\delta_{sd}^{LL}|$, much 
tighter constraints on $\mbox{Im}\,\delta_{sd}^{LL}$ can be 
derived by applying the severe bounds on the product 
$\delta_{sd}^{LL}\,\delta_{sd}^{RR}$ obtained from the 
chirally-enhanced contributions to $K$--$\bar K$ mixing. Even in 
this case, however, 
the supersymmetric contribution to $\epsilon'/\epsilon$ can still 
be of order $10^{-3}$.\cite{us} 

In summary, we have shown that in supersymmetric extensions of 
the Standard Model gluino box diagrams can yield a large 
$\Delta I=\frac32$ contribution to $\epsilon'/\epsilon$, which
only requires moderate mass splitting between the right-handed 
squarks. In a large region of parameter space, the measured value 
of $\epsilon'/\epsilon$ implies a significantly stronger bound 
on $\mbox{Im}\,\delta_{sd}^{LL}$ than is obtained from
$K$--$\bar K$ mixing.

\nonumsection{Acknowledgements}
\noindent
It is a pleasure to thank Alex Kagan for collaboration on the 
subject of this talk. This work was 
supported in part by the National Science Foundation.

\nonumsection{References}


\noindent

\begin{thebibliography}{99}

\bibitem{Gabb}
J. Hagelin, S. Kelley and T. Tanaka, \npb{415}{1994}{293};
E. Gabrielli, A. Masiero and L. Silvestrini, \plb{374}{1996}{80};
F. Gabbiani, E. Gabrielli, A. Masiero and L. Silvestrini, 
\npb{477}{1996}{321};
A.L Kagan, \Prd{51}{1995}{6196}.

\bibitem{SUSY}
A. Masiero and H. Murayama, \Prl{83}{1999}{907};
S. Khalil, T. Kobayashi, and A. Masiero, \Prd{60}{1999}{075003};
S. Khalil and T. Kobayashi, \plb{460}{1999}{341};
K.S. Babu, B. Dutta and R.N. Mohapatra, \Prd{61}{2000}{091701};
S. Baek, J.-H. Jang, P. Ko and J.H. Park, \Prd{62}{2000}{117701};
R. Barbieri, R. Contino and A. Strumia, \npb{578}{2000}{153};
A.J. Buras et al., \npb{566}{2000}{3};
G. Eyal, A. Masiero, Y. Nir and L. Silvestrini, 
\jhep{9911}{1999}{032}.

\bibitem{us}
A.L. Kagan and M. Neubert, \Prl{83}{1999}{4929}.

\bibitem{long}
M. Ciuchini et al., \jhep{9810}{1998}{008}.

\bibitem{Gian}
S. Dimopoulos and G.F. Giudice, \plb{357}{1995}{573}. 

\end{thebibliography}
\end{document}